\documentclass[12pt,letterpaper]{article}
\usepackage[affil-it]{authblk}
\usepackage{amsmath, amsfonts, amsthm, amssymb}
\usepackage[usenames,dvipsnames]{color}
\usepackage{comment}
\usepackage{ifpdf}
\usepackage{setspace}
\usepackage[utf8]{inputenc}
\usepackage[english]{babel}
\usepackage{graphicx}  
\usepackage{breqn}
\usepackage{mathrsfs}
\usepackage[left=.8in,right=.8in,top=.8in,bottom=.8in]{geometry}                  

\title{A Rotating Black Hole Solution for Shape Dynamics}
\author{ Henrique Gomes,  Gabriel Herczeg}
\affil{Department of Physics, University of California, Davis}
\affil{Davis, California 95616}
\date{\today}

\begin{document}

\maketitle
\begin{abstract}

Shape dynamics is a classical theory of gravity which agrees with general relativity in many important aspects, but which possesses different gauge symmetries and can present some fundamental global differences with respect to Einstein spacetimes. 
Here, we present a general procedure for (locally) mapping stationary, axisymmetric general relativity solutions onto their shape dynamic counterparts. 
We focus in particular on the rotating black hole solution for shape dynamics and show that many of the properties of the spherically symmetric solution are preserved in the extension to the axisymmetric case: it is also free of physical singularities, it does not form a space-time at the horizon, and it possesses an inversion symmetry about the horizon. 

\end{abstract}

\section{Introduction}

\subsection{Shape Dynamics}

Shape dynamics is a classical, Hamiltonian theory of gravity in which solutions are described by the time evolution of spatial (three-dimensional Riemannian) conformal geometries \cite{SD_first}. The canonical variables are given by a  Riemannian metric $g_{ij}$ and its conjugate momentum $\pi^{ij}$. It is instructive to compare shape dynamics with the ADM formulation of general relativity \cite{ADM}. In the ADM formalism, the gauge symmetries are spatial diffeomorphism and (on--shell) refoliation invariance. Refoliation invariance allows one to transform from a solution on one family of spacelike hypersurfaces to a physically equivalent solution on another. Shape dynamics does not possess refoliation invariance; the solutions are described by the evolution of a conformal class of Riemannian geometries. 

Rather than refoliation invariance, the equations of motion of shape dynamics are invariant under local scale transformations of the spatial metric. The result is a theory which possesses different gauge symmetries than general relativity, but which is nevertheless generically \emph{locally} equivalent to it. Local equivalence holds in the sense that around a generic point in a solution to Einstein's equations there is a local patch which can  be directly mapped onto a shape dynamics solution, and vice-versa. The mapping is achieved by simultaneously applying (partial) gauge fixings on each theory \cite{Linking}.  

Local scale transformations (here called spatial Weyl transformations) are generated by the trace of the momentum $\pi = \pi^{ij} g_{ij}$. The statement that spatial Weyl transformations are a gauge symmetry of shape dynamics\footnote{The full set of Weyl transformations are easier to implement in case one has a  non-closed spatial manifold. In case the manifold is closed, shape dyanamics is constructed with the group of Weyl transformations that preserve the total volume of space.} is translated in the Dirac constraint formalism  into $\pi = 0$ being a first class constraint on the phase space of shape dynamics. It is only for solutions of ADM for which one can find a global foliation in maximal slicing  -- i.e. that for which $\pi=0$ on each hypersurface-- that ADM can be dual to shape dynamics. For certain important cases, such as the one we will investigate here, this is not the case. 

{For this reason we carefully distinguish between the notion of line-element and space-time metric. This is because shape dynamics does not possess space-time diffeomorphism symmetry, as we have stressed above. A reconstructed solution always exists and gives rise to a line element in the maximal foliation, but in many circumstances the line element will not form a non-degenerate, or a non-singular, space-time metric.} 


\subsection{Black Hole Solutions}\label{BHS}

Despite the local equivalence of shape dynamics and general relativity, there nevertheless exists the possibility that corresponding solutions of shape dynamics and general relativity may have different global structures. In particular, it was recently shown in \cite{Birkhoff} that while shape dynamics possesses a unique asymptotically flat, spherically symmetric solution that agrees with general relativity near spatial infinity, the solution differs physically from the Schwarzschild solution inside and at the event horizon. The purpose of this paper is to generalize these results to the case of stationary, axisymmetric solutions, and to explicitly derive the rotating black hole solution for shape dynamics. We will discuss how the solutions are related to the corresponding exterior general relativity solutions, and how they depart from one another at the horizon. Finally, we will make contact with the spherically symmetric solution by considering the zero angular momentum limit of the rotating solution.

Before considering the axisymmetric case, it is instructive to review the main results of \cite{Birkhoff} for the spherically symmetric solution of shape dynamics. 

\subsubsection*{The isotropic line element}
The Birkhoff theorem for general relativity states that the only vacuum solution that is spherically symmetric is in the isometry class of the Schwarzschild line element. For shape dynamics, an analogous theorem exists, but there one must also provide boundary conditions on the dynamical variables. By reconstructing the line element we obtain: 
\begin{equation}
 \label{equ:SD_new}
ds^2= -\left(\frac{1-\frac{m}{2r}}{1+\frac{m}{2r}}\right)^2dt^2+ \left(1+\frac{m}{2r}\right)^4\left(dr^2+r^2(d\theta^2+\sin^2\theta d\phi^2)\right)
\end{equation}
where $\phi$ and $\theta$ are the usual angular coordinates. The radial coordinate $r$ is related to the Schwarzschild radial coordinate,  $ r_{\mbox{\tiny{s}}}$  only for $r>m/2$. In that regime the line elements are isometric, and the relation between the two radial coordinates is given by
\begin{equation}\label{Schwarz to Iso}
r_{\mbox{\tiny{s}}} = r\left(1 + \frac{m}{2r}\right)^2
\end{equation}
At $r =m/2$ the coordinate change breaks down. The  line element \eqref{equ:SD_new} is degenerate at the horizon, where the 4-volume of the reconstructed metric collapses. The collapse is a coordinate-independent statement in the space-time view, as the vanishing of the metric determinant $\det{(g_{\mu\nu})}=0$ is a physical effect.  This defect in the space-time view, however, does not afflict a shape dynamics interpretation, since shape dynamics does not require the emergence of space-time for a solution to be well-defined. 

An observer in the asymptotically flat region will not see any difference between this solution and a Schwarzschild black hole at the classical level. Nonetheless, the fact that there is indeed a \emph{physical} difference can be seen directly by observing differences in physical statements made about general relativistic vacuum spacetime solutions. 
For instance, a simple calculation shows that the infalling radial geodesic takes infinite proper time to reach $r=0$ \cite{Poplawski}. This is of course a physical, observable distinction between the spherically symmetric shape dynamics solution and the Schwarzschild solution in general relativity. 
In summary, the line element in question, forms a bona-fide vacuum wormhole solution. For general relativity it is well known that to form such solutions one must include a singular unstable thin-shell of matter at the horizon \cite{Visser_book}. 

Furthermore, the line element \eqref{equ:SD_new} can be shown to form from a thin-shell collapse \cite{Tim_collapse}. This is possible because from the shape dynamics perspective it is \emph{reduced phase space}  continuity of the solutions that is fundamental, not space-time continuity. If one demands space-time continuity, one recovers the usual Schwarzschild solution, if one demands reduced phase space continuity of the solution, one recovers a collapse asymptoting  to the line element \eqref{equ:SD_new}. 
 
Another interesting property that we will also find in the axisymmetric case is that the transformation $r\to m^2/(4r)$  can be checked to leave the form of the line element (\ref{equ:SD_new}) invariant. Thus the solution has the property of \emph{inversion}, associated to conformal invariance (as for instance in the method of images in electrodynamics). The presence of this inversion symmetry, along with asymptotic flatness, is enough to see that there cannot be a physical singularity at $r = 0$ in the shape dynamic solution.

It is natural to ask if this singularity-avoiding property persists in all solutions of shape dynamics. We should mention that space-times in maximal slicing have a well-known singularity avoidance property for its Eulerian observers \cite{Gourgoulhon}. Due to the close relationship between shape dynamics and ADM in maximal slicing, there is good reason to think that shape dynamics might share this singularity-avoiding property.  One of the purposes of this paper is to show that at least in the case of axisymmetric solutions this is indeed the case. 

\section{Stationary, Axisymmetric Solutions}\label{SAS}

We will consider the stationary axisymmetric line element:
\begin{equation}\label{Axi}
ds^2 = -(N^2 - \Omega\Psi\xi^2)dt^2 + \Omega[(dx^1)^2 + (dx^2)^2 + \Psi d \phi^2] + 2\Omega\Psi\xi d \phi dt
\end{equation}
where $N$, $\xi$ and $\Omega$ are the lapse, shift and conformal factor respectively, $\Psi$ is a function that determines the entire spatial conformal geometry, and all functions depend exclusively on $x^1$ and $x^2$. Strictly speaking $\xi ^a=\xi\delta^a_{\phi}$ is the shift vector but we will from time to time abuse language and refer to the scalar $\xi$ simply as ``the shift" since it is often more convenient to work directly with this quantity.

We would like to show that the line element \eqref{Axi} is already in maximal slicing, i.e., that the momentum conjugate to the spatial metric has vanishing trace, $\pi = 0$. 
To this end, consider Hamilton's equation for the time derivative of the spatial metric, which takes the form
\begin{equation}\label{Hamilton}
\dot{g}_{ij} = 2N(\Omega^3\Psi)^{-1/2}(\pi_{ij}-\frac{1}{2}\pi g_{ij}) + \mathcal{L}_{\xi}g_{ij}
\end{equation}     
where $\mathcal{L}_{\xi}g_{ij}$ denotes the Lie derivative of the spatial metric along the shift vector. Since the spatial metric is independent of $\phi$ we have
\begin{dmath}\label{lie metric}
\mathcal{L}_{\xi}g_{ij}  =   \partial_i \xi g_{\phi j} + \partial_j \xi g_{i \phi}  =   
2\Omega\Psi{\delta^{\phi}}_{(i}\xi_{,j)} 
\end{dmath}
where we have introduced the round parentheses for symmetrization of indices and the comma for coordinate derivatives.   Putting \eqref{lie metric} into \eqref{Hamilton}, and noting that the spatial metric is independent of $t$, we obtain
\begin{equation}\label{Hamilton2}
2N(\Omega^3\Psi)^{-1/2}(\pi_{ij}-\frac{1}{2}\pi g_{ij}) + 
2 \Omega\Psi {\delta^{\phi}}_{(i}\xi_{,j)}= 0
\end{equation}
Contracting with $g^{ij}$ yields: 
\begin{eqnarray}\label{TraceHam}
2N(\Omega^3\Psi)^{-1/2}(\pi - \frac{3}{2}\pi) +
2\xi_{,i}g^{i\phi} &=& \nonumber \\ 
-N(\Omega^3\Psi)^{-1/2}\pi +2(\Omega\Psi)^{-1}\xi_{,\phi} &=& 0 
\end{eqnarray}
Noting that $\xi$ is independent of $\phi$, we find that 
\begin{equation}\label{maximal}
-N(\Omega^3\Psi)^{-1/2}\pi = 0. 
\end{equation}
Thus we find that whenever a general relativity solution can be written as \eqref{Axi}, either it is maximally sliced or $N(\Omega^3\Psi)^{-1/2}=0$. We will assume that $\Omega$ and $\Psi$ are bounded (have finite values on compact sets). If furthermore $N$ can vanish only on singular subsets of $M$, then it follows that $\pi=0$ everywhere except at most a singular subset of $M$. Continuity of $\pi$ then demands that it is zero everywhere in space. 

Furthermore, a general axisymmetric solution of Einstein's equations can be put into the form \eqref{Axi} \cite{Papapetrou}, and thus in principle the line element can be formed by a reconstruction of a shape dynamics solution. We will see in section \ref{completeness} that indeed there are no obstructions on the shape dynamics side of the duality.  

\section{Rotating Black Hole Solutions}\label{RBHS}

\subsection{The Solution}

The Kerr space-time is a stationary, axisymmetric solution to Einstein's equations. As such, it must be possible (at least locally) to cast it  in the form \eqref{Axi}. Indeed, it was shown in \cite{Papapetrou} that the Kerr metric can be put in the form 
\begin{equation}\label{Kerr PS} 
ds^2 = -\lambda^{-1}(dt - \omega d\phi)^2 + \lambda [m^2e^{2\gamma}(d\mu^2 + d\theta^2) + s^2d\phi^2]
\end{equation}
where 
\begin{eqnarray}\label{Kerr functions PS} 
s \hspace{5pt} &=& mp\sinh\mu\sin\theta \nonumber \\
e^{2\gamma} &=& p^2\cosh^2\mu  + q^2\cos^2\theta - 1 \\
\omega \hspace{5pt} &=& e^{-2\gamma}\left[2a\sin^2\theta(p\cosh\mu + 1)\right] \nonumber \\
\lambda \hspace{5pt} &=& e^{-2\gamma}\left[(p\cosh\mu + 1)^2 + q^2\cos^2\theta \right] \nonumber \\
\end{eqnarray} 
where $p = \sqrt{1-a^2/m^2}$, $q = a/m$,  $a = J/m$, and where $m$ and $J$ are the mass and angular momentum.  It is easy to see that the metric written in terms of these (prolate spheroidal) coordinates is in the form \eqref{Axi}, so it can be mapped directly onto a shape dynamics solution via ADM decomposition. The lapse and shift can be read off of the line element:
\begin{equation}\label{lapse shift}
 N^2 = \lambda^{-1}\left(\frac{\omega^2}{\lambda^2s^2-\omega^2} + 1\right), \hspace{10pt} \xi = \frac{\omega}{\lambda^2-\omega^2}.
\end{equation}
 Putting \eqref{lapse shift} into \eqref{Hamilton2} and solving for $\pi_{ij}$, we find
\begin{equation}\label{first pi}
\pi_{ij} = -\left(\frac{\Omega^5\Psi^3}{N^2}\right)^{1/2}\left[{\delta^{\phi}}_{(i}{\delta^{\mu}}_{j)}\xi_{,\mu} +{\delta^{\phi}}_{(i}{\delta^{\theta}}_{j)}\xi_{,\theta} \right].
\end{equation}
It is easy to see that at the horizon, where $s = 0$, the lapse goes to zero. One might then worry that $\pi_{ij}$ might diverge there, violating phase space continuity. However, we should note that $\Psi$ goes as $s^2$, so clearly the prefactor in the  \eqref{first pi} goes to zero at the horizon.

Since it is more familiar, we would like to show that \eqref{Kerr PS} is locally diffeomorphic to the Kerr metric written in Boyer-Lindquist coordinates. The coordinate transformation
\begin{equation}\label{PS to BL}
\mu = \cosh^{-1}\left(\frac{r_{\mbox{\tiny{BL}}} - m}{\sqrt{m^2 - a^2}}\right)
\end{equation}
brings \eqref{Kerr PS} into the desired form:
\begin{equation}\label{Kerr BL}
ds^2 = -\frac{\Delta}{\Sigma}\left(dt - a \sin^2\theta d\phi\right)^2 + \frac{\sin^2\theta}{\Sigma}\left((r_{\mbox{\tiny{BL}}}^2 + a^2)d\phi - a dt\right)^2 + \frac{\Sigma}{\Delta}dr_{\mbox{\tiny{BL}}}^2 + \Sigma d\theta^2
\end{equation} 
where 
\begin{eqnarray}\label{Kerr functions BL}
\Delta  &=&  r_{\mbox{\tiny{BL}}}^2 - 2mr_{\mbox{\tiny{BL}}} + a^2 \nonumber \\
\Sigma   &=& r_{\mbox{\tiny{BL}}}^2 + a^2 \cos^2\theta \\
\end{eqnarray}

 It is interesting to note that the change of coordinates \eqref{PS to BL} is purely spatial, so it would seem that the two forms are equally valid from the point of view of shape dynamics. This is not the case, however, since the transformation fails to be differentiable at the event horizon, which is conveniently labeled in prolate spheroidal coordinates by $\mu = 0$. The transformation is therefore not globally a diffeomorphism, which is compatible with the fact that the ADM decomposition of the Kerr metric in Boyer-Lindquist coordinates constitutes a shape dynamics solution only outside the event horizon. The solution written in prolate spheroidal coordinates possesses no such deficiency, and represents a complete solution of the shape dynamics equations of motion, even though it too breaks down at the event horizon when viewed from the perspective of general relativity, as we will see in following section. 

\subsection{Completeness of the Solution}\label{completeness}

It is fairly easy to see that the Kerr metric written in prolate spheroidal coordinates breaks down at the event horizon from the point of view of general relativity. To make this breakdown explicit, we need only consider the determinant of the space-time metric.
									
\begin{equation}\label{det(g4)}
\det(g^{(4)}) 
= -\lambda^2m^4e^{4\gamma} s^2.
\end{equation}
Clearly, $s$ goes to zero at $\mu = 0$ while $\lambda$ and $e^{4\gamma}$ remain finite, indicating that $\det(g^{(4)})$ goes to zero at $\mu = 0$ and consequently that the space-time metric is noninvertible there. This shows that the Kerr metric written in terms of prolate spheroidal coordinates posesses a coordinate singularity at the event horizon, $\mu = 0$. This is then not a complete solution from the point of view of general relativity, but must be regarded as a solution only in the region outside of the event horizon. This is not the case, however, from the point of view of shape dynamics, where the conformal spatial geometry, rather than the space-time geometry, is considered fundemental. It can be immediately seen that the determinant of the spatial metric is given by $\det(g) = m^4e^{4\gamma}\lambda^2(\lambda s^2-\lambda^{-1}\omega^2) \neq 0$ for all real values of $\mu$, and diverges only as $\mu \to \pm\infty$. This represents a rather dramatic departure from the general relativistic solution which requires an entirely different interior to the event horizon, possessing well known technical problems such as physical singularities and closed timelike curves. 

\subsection{Shape Dynamic Horizons: Classical Firewalls?}\label{horizons}

It is interesting to note that while the breakdown in space-time geometry that occurs at the horizon is not forbidden by shape dynamics (in fact, it is \textit{required}), it does have some interesting consequences for infalling observers. The well known ``no drama" result of general relativity does not hold in shape dynamics because the equivalence principle is an emergent property of shape dynamics, not an axiom, and this property fails to emerge precisely at the event horizon. To better understand the nature of the horizon, consider the following argument.

Let us assume that in the interior and exterior regions, the trajectories of observers are described by (conformal equivalence classes of) timelike geodesics of the reconstructed line element. Since there are no outgoing timelike geodesics in the exterior region that originate in the interior region, mirror symmetry\footnote{We assume that a generic stationary shape dynamic black hole will possess inversion, or mirror symmetry about the horizon. Over-extreme black holes, which do not possess horizons are exempt from these considerations. There is some evidence to suggest that this assumption may be violated for boundary conditions other than asymptotic flatness, but as of yet no such solutions have been found. We will show in section \ref{ZAML} that the rotating shape dynamic black hole does possess mirror symmetry.} seems to demand that there are no ingoing timelike geodesics originating in the exterior region. It would appear then, that the horizon must be interpeted as the location where timelike geodesics terminate in both regions. This is not the case, however, as can be seen by noting that the lapse goes to zero at the horizon, and becomes negative in the interior region. As a result, the timelike geodesics in the interior region should be interpreted in a time-reversed fashion \cite{Poplawski}. Only in this peculiar manner can the ingoing geodesics in the exterior region  be smoothly connected to ingoing geodesics in the interior region.

The consequences of the argument presented above can be significant for infalling observers. Rather than passing through the horizon uneventfully, infalling observers might be able to perform a measurement to determine the instant at which they pass into the time-reversed parallel universe. Indeed, it can be shown that the expansion scalar of congruence of time-like geodesics suffers a finite discontinuity, changing signs at the horizon \cite{Poplawski}. This might provide infalling observers with a well-defined signal of having crossed the horizon. A comoving ball of matter as measured by an infalling observer will decrease in volume up to the horizon, at which point it will ``bounce" outward and begin to expand. One might argue that since the volume element is considered to be pure gauge in shape dynamics, the volume of a ball of matter is not a true physical observable. Indeed, this is partially true even in GR, since measurements of volume can be changed by for example performing a local lorentz transformation. However, since there is a finite discontinuity in the expansion scalar, there should be no continuous spatial diffeomorphism or conformal transformation that will remove this discontinuity. Hence, while different observers might disagree about the details of the measurement, it is possible that  all observe the volume bounce. This matter should be further investigated in the context of the infalling shell of dust. 

Since the shape dynamic description of stationary black holes requires a violation of the equivalence principle at the event horizon, it is an exciting possibility that quantum shape dynamic black holes may change the picture of the firewall paradox  \cite{AMPS}. It is too early to tell at this writing what, if any, insights shape dynamics can contribute to this debate, but the authors are currently investigating the properties of shape dynamic horizons in this context.

\section{Alternative Gauge-Fixing}\label{AGF}
\subsection{Equations of Motion}
The spatial conformal invariance of shape dynamics allows us to cast the solution in an alternative form by extracting a common scalar function from the spatial metric.  The transformed metric and conformal factor are
\begin{eqnarray}\label{new gauge}
g_{ij} &=& m^2\left(\delta^{\mu}_i \delta^{\mu}_j + \delta^{\theta}_i \delta^{\theta}_j + m^{-2}e^{-2\gamma}(s^2 - \lambda^{-2}\omega^2)\delta^{\phi}_i \delta^{\phi}_j\right) \nonumber \\ 
&=&  m^2\left(\delta^{\mu}_i \delta^{\mu}_j + \delta^{\theta}_i \delta^{\theta}_j + \Psi\delta^{\phi}_i \delta^{\phi}_j\right) \\
\Omega &=& e^{-2\gamma} \lambda^{-1} = [(p\cosh\mu + 1)^2 + q^2\cos^2\theta]^{-1} \nonumber 
\end{eqnarray} 
We know from the gauge symmetries of shape dynamics that the transformed solution must also be a solution to the shape dynamics equations of motion. For the spatially noncompact case, the equations of motion read
\begin{equation}\label{gDot}
\dot{g}_{ij} = 4\rho g_{ij} + 2e^{-6\Phi}\frac{N}{\sqrt{g}}\pi_{ij} + \mathcal{L}_{\xi}g_{ij}  \\
\end{equation}

\begin{eqnarray}\label{piDot}
\dot{\pi}^{ij} &=& N e^{2\Phi}\sqrt{g}(R^{ij} - 2 \Phi^{;ij} + 4 \Phi^{;i}\Phi^{;j} - \frac{1}{2}Rg^{ij} + 2\nabla^2\Phi g^{ij}) \nonumber  \\
 & & -e^{2\Phi}\sqrt{g}(N^{;ij} - 4\Phi^{(,i}N^{,j)} - \nabla^2 Ng^{ij}) + \mathcal{L}_{\xi}\pi^{ij} - 4\rho\pi^{ij}  \\
 & & -\frac{N}{\sqrt{g}}e^{-6\Phi}(2\pi^{ik}\pi^{j}_{k} - \pi^{kl}\pi_{kl}g^{ij}) \nonumber \\ \nonumber
\end{eqnarray}
 \noindent where $\rho$ is a lagrange multiplier associated with the conformal constraint, and $\Phi = \ln\Omega$ satisfies the Lichnerowicz-York  equation:
\begin{equation}\label{LY}
\nabla^2\Omega + \frac{R}{8}\Omega - \frac{1}{8}\pi^{ij}\pi_{ij}\Omega^{-7} = 0 
\end{equation}

We can eliminate $\rho$ by putting \eqref{new gauge} into \eqref{gDot}, taking the trace, and requiring that $\pi = 0$, which immediately yields $\rho = 0$. Putting this back into \eqref{gDot}, we find that 
\begin{eqnarray}\label{momenta}
\pi_{ij} &=& \frac{\Psi^{3/2}e^{6\Phi}}{m^2N}\left( \xi_{,\mu}\delta^{\mu}_{(i}\delta^{\phi}_{j)} + \xi_{,\theta}\delta^{\theta}_{(i}\delta^{\phi}_{j)}\right) \nonumber \\
\pi^{ij} &=& \frac{\Psi^{1/2}e^{6\Phi}}{m^2N}\left(\xi_{,\mu}\delta^{(i}_{\mu}\delta^{j)}_{\phi} + \xi_{,\theta}\delta^{(i}_{\theta}\delta^{j)}_{\phi}\right)
\end{eqnarray}

While the equations of motion are somewhat more complicated in this gauge, it can be shown that the transformed solutions \eqref{new gauge}, \eqref{momenta} do indeed satisfy \eqref{gDot}, \eqref{piDot}. One key advantage of this alternative gauge fixing is that the spatial metric now possesses only one functional degree of freedom, $\Psi$. Now the entire spatial geometry can be expressed in terms of $\Psi$ and it's derivatives alone. This simplified form of the metric can be exploited for the purposes of analyzing the spatial conformal structure of the solution. In particular, it wll aid us in searching for singularities in the conformal structure.

\subsection{Conformal Regularity of the Horizon}
The simplified form of the metric arising from our change of conformal gauge produces a correspondingly simplified connection and curvature tensor. The interested reader is refered to the appendix for the calculation of these quantities. We define the Cotton tensor by

\begin{equation}\label{Cotton1}
\mathscr{C}_{ijk} := \nabla_k\left(R_{ij} - \frac{1}{4}Rg_{ij}\right) - \nabla_j\left(R_{ik} - \frac{1}{4}Rg_{ik}\right)
\end{equation}

\noindent where $\nabla$ denotes covariant differentiation with respect to the spatial metric. The rank-two Cotton-York tensor, $C^{ij}$, can be defined by it's relation to the Cotton tensor:

\begin{equation}\label{Cotton}
C^{ij} := -\frac{1}{2}g^{mj}\epsilon^{ikl}\mathscr{C}_{mkl}
\end{equation}

The Cotton tensor contains all of the information on the conformal geometry of a three-dimensional Riemannian manifold \cite{Gourgoulhon} in much the same way that the Weyl tensor (which vanishes identically in three dimensions) captures information on the conformal geometry in higher dimensions. Like the Weyl tensor in higher dimensions, the Cotton tensor is completely traceless, conformally invariant, and vanishes if and only if the manifold is conformally flat.  From \eqref{Cotton-York}, \eqref{CY squared}, and \eqref{Cotton} we see that if $C^2 := C^{ij}C_{ij}$ diverges then there must be a singularity in the Cotton tensor. A singularity in the Cotton tensor would signal the presence of a breakdown of the conformal geometry, i.e it would be a \textit{physical} singularity from the perspective of shape dynamics. It is therefore useful to show that $C^2$ is finite as a heuristic argument that no such physical singularities are present. In this sense, although $C^2$ is not strictly speaking conformally invariant (it transforms as $C^2 \to \Omega^{5/2}C^2$  under $g_{ij} \to \Omega g_{ij}$), it can be thought of in anology with the Kretschmann invariant in general relativity. Moreover, if we assume that $\Omega$ is bounded in the sense described in section \ref{SAS}, then the presence of conformal covariance as opposed to conformal invariance is essentially irrelevant for the purposes of identifying singularities in the conformal structure.  

Intuition suggests that the points we should scrutinize most carefully are the horizon and the limit as $\mu \to -\infty$, since we have already noted some peculiaraties about the former, and the latter seems analogous to the singularity in general relativity. We should note, however, that the latter is really just spatial infinity, so asymptotic flatness ensures that there are no conformal singularities there. 

 To help simplify the calculation for the horizon, we can note that since $\Psi$ is an even function of $\mu$, any odd number of $\mu$ derivatives acting on $\Psi$ will be zero when evaluated on the horizon. Taking this into account we can put \eqref{CY squared} in the simplified form
\begin{equation}\label{CY squared on horizon}
C^2(0, \theta) =  \left.\left[ \frac{1}{4\Psi^2}\Psi_{,\theta}\Psi_{,\theta\theta} - \frac{1}{4\Psi}\left( \Psi_{,\mu\mu\theta} + \Psi_{,\theta\theta\theta} \right) \right] \right|_{\mu = 0}
\end{equation}

\noindent At the horizon, we have 

\begin{equation}\label{psi on horizon}
\Psi(0, \theta) = \frac{4a^2\sin\theta}{m^2\left(4 + q^2\cos^2\theta\right)^2}
\end{equation}

\noindent which is nonzero except on the axis of rotation $\theta = \{0, \pi\}$. We will therefore have to carefully analyze the limits as we approach these points. A lengthy but straightforward calculation yields the other ingredients of \eqref{CY squared on horizon}:

\begin{align}\label{ingredients}
\Psi_{,\theta}(0,\theta) &= -\frac{2^8 a^2 q^2 \sin^2\theta\cos\theta}{m^2(4+q^2\cos^2\theta)^3} \nonumber \\
\Psi_{,\theta\theta}(0,\theta) &= \frac{2^8}{m^2(4+q^2\cos^2\theta)^3}\left[ a^2q^2\sin^2\theta(\sin^2\theta-\cos^2\theta) - \frac{6a^2q^4\sin^3\theta\cos^2\theta}{4+q^2\cos^2\theta} \right] \nonumber \\
\Psi_{,\theta\theta\theta}(0,\theta) &=  \frac{2^9a^2q^2\sin^2\theta\cos\theta}{m^2(4+q^2\cos^2\theta)^4}\left[ 3\cdot2^3q^4\sin^2\theta\cos^2\theta + 3^2q^2(\sin^2\theta-\cos^2\theta) + 2(4+q^2\cos^2\theta) \right] \nonumber \\
\Psi_{,\theta\mu\mu}(0,\theta) &= 4\sin\theta\cos\theta + 3\cdot2^{10}\frac{a^2q^2}{m^2}\frac{\sin\theta\cos\theta}{(4+q^2\cos^2\theta)^4}
\end{align}

Clearly, none of the functions in \eqref{ingredients} can diverge for any values of $\theta$. Moreover, if we insert \eqref{ingredients} back into \eqref{CY squared on horizon}, we see that $C^2 = 0$ at $\theta = 0$ and $\theta = \pi$. So despite the peculiar behaviour of the horizon when viewed from the four-dimensional perspective, we do not see any conformal singularities manifesting themselves in the Cotton-York tensor at the horizon.  

\section{Zero Angular Momentum Limit}\label{ZAML}

Finally, we wish to demonstrate that in the zero angular momentum limit $a = 0$ of the solution presented in section \ref{RBHS}, we recover the spherically symmetric solution \eqref{equ:SD_new} presented in \cite{Birkhoff}. From the definitions of $p$ and $q$, we can see that $a = 0$ implies $p = 1$, $q = 0$. Putting these limits into \eqref{Kerr functions PS} we obtain

\begin{align}\label{PS a = 0}
&s \hspace{7.7pt}= \hspace{7pt} m\sinh\mu\cos\theta \nonumber \\
&\lambda \hspace{7pt} = \hspace{7pt} \frac{(\cosh\mu + 1)^2}{\sinh^2\mu}  \\
&\omega \hspace{7pt} = \hspace{7pt} 0 \nonumber \\
&e^{2\gamma} = \hspace{7pt} \sinh^2\mu \nonumber
\end{align}  

Inserting \eqref{PS a = 0} into \eqref{Kerr PS} gives the spherically symmetric line element written in terms of prolate spheroidal coordinates.

\begin{equation}\label{Kerr PS a = 0}
ds^2 =  -\frac{\sinh^2\mu}{(\cosh\mu + 1)^2}dt^2 +  m^2(\cosh\mu + 1)^2\left( d\mu^2 + d\theta^2 + \sin^2\theta d\phi^2\right)
\end{equation}

One can already see the isotropic character of the solution as written in terms of prolate spheroidal coordinates. In order to obtain the form of the solution presented in \cite{Birkhoff} we perform the spatial diffeomorphism $r = \frac{m}{2}e^{\mu}$ which can be checked to reproduce \eqref{equ:SD_new}, as desired. It is important to note that unlike \eqref{PS to BL}, this transformation and its inverse are differentiable  everwhere, so it is a global diffeomorphism--i.e. the transformation is pure gauge. It is interesting to note that the inversion symmetry of the spherically symmetric solution takes on a simplified form when written in terms of prolate spheroidal coordinates. In this case the inversion symmetry is manifested by the fact that the line element is an even function of $\mu$. Indeed, this is the case even before we take the limit $a = 0$, so we conclude that the axisymmetric solution also possesses an inversion symmetry under $\mu \to -\mu$. This is a nice representation of the symmetry since it emphasizes that what we are doing is \emph{reflecting} about the event horizon $\mu = 0$ into the corresponding point of the time-reversed mirror universe.

\section{Discussion}
We have provided the most general local form of stationary, axisymmetric vacuum solutions to the shape dynamics equations of motion and used this result to obtain the rotating black hole solution for shape dynamics. The rotating black hole solution preserves many of the striking features of the spherically symmetric case. It possesses a powerful inversion symmetry about the horizon where it does not form a space-time, and it seems to completely avoid physical singularities. The inversion symmetry and singularity avoidance are perhaps even more surprising in the rotating case, since the corresponding general relativity solution is so complicated in the interior region, possessing a ringlike physical singularity, closed timelike curves and an inner cauchy horizon. The shape dynamics solution, by contrast, avoids all of these difficulties by creating at the event horizon the time-reversed mirror universe that allows the matter source to expand out to an inner spatial infinity and avoid collapsing to a singularity. 

The extreme and over-extreme Kerr solutions can be mapped onto their shape dynamic counterparts using the same arguments presented above. These solutions are also presented in \cite{Papapetrou} in the form \eqref{Axi}, making the mapping to shape dynamics almost trivial. In the overextreme case, the four dimensional line element associated with the shape dynamics solution is globally related to Boyer-Lindquist coordinates by a spatial diffeomorphism. It is not yet clear whether the naked singularity persists in the shape solutions--it is likely that this singularity is fundementally four dimensional in nature and does not appear in shape dynamics. If it does persist in shape dynamics it would be the first singular solution to shape dynamics of which the authors are aware at this writing.

Probably the most exciting feature of the black hole solutions for shape dynamics is that they do not form a space-time at the horizon. Susskind and Maldacena have argued \cite{Cool} that there is close relationship between entanglement and (non-traversable) wormholes in the context of a possible resolution to the firewall paradox. In shape dynamics, the stationary black hole solutions seem to be traversable wormholes\footnote{A more definitive answer to the question of traversability can only be answered by coupling matter degrees of freedom. This has been done in the spherically symmetric case for a collapsing thin shell of matter \cite{Tim_collapse}. } that suggest two sources for a possible resolution to this debate: black holes are correctly described by shape dynamics, and there is no paradox because the equivalence principle breaks down at the horizon and/or there is no singularity. The obvious next steps in these considerations are to look at the semiclassical behavior of these solutions, to analyze their thermodynamic properties, and to consider the behavior of quantum fields in the presence of a stationary shape dynamic black hole background. 

The fact that the solution is well-behaved at the horizon gives it an advantage over similar traversable wormhole models in general relativity, which are generically unstable. Furthermore, since the latter require a delta function contribution to the curvature scalar at the horizon \cite{Visser_book}, these solutions must be regarded as singular space-times. Moreover, if traversable wormholes admit a more consistent quantum mechanical interpretation than the standard stationary black hole solutions in general relativity, then since these solutions arise naturally in shape dynamics\footnote{It should be noted that these seem to be the \emph{only} stationary black hole solutions that arise naturally in shape dynamics. The singularity avoidance theorems for Eulerian observers in maximal slicing make it implausible that anything like ordinary Schwarzschild or Kerr solutions could be made to satisfy the shape dynamics equations of motion everywhere.} this might be a hint that shape dynamics is a more consistent classical theory of gravity than general relativity for the purposes of quantization. As a last remark, let us mention that since the shape dynamics  solutions are indistinguishable from the corresponding general relativity solutions in the asymptotically flat region, they are on equally solid ground from the point of view of current empirical astrophysical observations.

\subsection*{Acknowledgments}
We would like to thank Steven Carlip for his support and for his many helpful questions, comments and constructive criticisms. We would also like to thank Vasudev Shyam and Joshua Cooperman for their valuable input and insightful questions about this and related research. HG was supported in part by the U.S.Department of Energy under grant DE-FG02-91ER40674.

\subsection*{Appendix}

The non-zero connection coefficients associated with the metric \eqref{new gauge} are given by

\begin{eqnarray}\label{connection}
\Gamma^{\phi}_{\mu\phi} &=& \frac{1}{2}(\ln\Psi)_{,\mu} \nonumber \hspace{32.27 pt}
\Gamma^{\phi}_{\theta\phi} = \frac{1}{2}(\ln\Psi)_{,\theta} \nonumber \\
\Gamma^{\mu}_{\phi\phi} &=& -\frac{1}{2}\Psi_{,\mu} \hspace{44.27 pt} 
\Gamma^{\theta}_{\phi\phi} = -\frac{1}{2}\Psi_{,\theta} 
\end{eqnarray}

From which we obtain the components of the  Ricci tensor

\begin{eqnarray}\label{Ricci}
R_{\mu\mu} &=& -\frac{1}{2}(\ln\Psi)_{,\mu\mu} - \frac{1}{4}[(\ln\Psi)_{,\mu}]^2 \nonumber \\
R_{\theta\theta} &=& -\frac{1}{2}(\ln\Psi)_{,\theta\theta} - \frac{1}{4}[(\ln\Psi)_{,\theta}]^2 \nonumber \\
R_{\mu\theta} &=& -\frac{1}{2}(\ln\Psi)_{,\mu\theta} - \frac{1}{4}(\ln\Psi)_{,\mu}(\ln\Psi)_{,\theta} \\
R_{\phi\phi} &=& -\frac{1}{2}(\Psi_{,\mu\mu} +\Psi_{,\theta\theta}) + \frac{1}{4\Psi}[(\Psi_{,\mu})^2 + (\Psi_{,\theta})^2] \nonumber \\
R_{\mu\phi} &=& R_{\theta\phi} = 0 \nonumber \\ \nonumber
\end{eqnarray}

and the Ricci scalar 

\begin{equation}\label{R}
R = \frac{1}{2\Psi^2}[(\Psi_{,\mu})^2 + (\Psi_{,\theta})^2] - \frac{1}{\Psi}(\Psi_{,\mu\mu} + \Psi_{,\theta\theta}) \bigskip \\
\end{equation}

Using (\ref{connection}), (\ref{Ricci}), and (\ref{R}) we can construct the Cotton-York tensor

\begin{equation}\label{Cotton-York}
C^{ij} = \epsilon^{ikl}\left(R^j_{\hphantom{j}l;k} - \frac{1}{4}\delta^j_l R_{,k} \right)
\end{equation}

which is by construction symmetric, traceless and transverse. Putting (\ref{connection}), (\ref{Ricci}), and (\ref{R}) into (\ref{Cotton-York}) yield the components of the Cotton-York tensor 

\begin{eqnarray}\label{CY components}
C^{\mu\mu} &=& C^{\mu\theta} = C^{\theta\theta} = C^{\phi\phi} = 0 \nonumber \\
C^{\mu\phi}  &=& \frac{1}{\Psi}R_{\phi\phi,\theta} -\frac{1}{4}R_{,\theta} \\
C^ {\theta\phi} &=& -\frac{1}{\Psi}R_{\phi\phi,\mu} + \frac{1}{4}R_{,\mu} \nonumber
\end{eqnarray}

from which we can form the scalar density 

\begin{eqnarray}\label{CY squared}
C^2 := C^{ij}C_{ij} &=& 2\Psi\left[ (C^{\mu\phi})^2 + (C^{\theta\phi})^2 \right] \nonumber \\ &=& \frac{1}{4\Psi^2}\left[ \Psi_{,\mu}\Psi_{,\mu\mu} + \Psi_{,\theta}\Psi_{,\theta\theta} + (\Psi_{,\mu} + \Psi_{,\theta})\Psi_{,\mu\theta} \right] \nonumber \\ 
&\hphantom{=}& -\frac{1}{4\Psi}\left( \Psi_{,\mu\mu\mu} + \Psi_{,\mu\mu\theta} + \Psi_{,\mu\theta\theta} + \Psi_{,\theta\theta\theta} \right).
\end{eqnarray}

\end{document}